\documentclass[prd,nofootinbib,amsfonts,notitlepage]{revtex4-1}
\usepackage{hyperref}
\usepackage{graphicx,bm,amsmath,color}
\usepackage[latin1]{inputenc}
\usepackage{bbold}
\usepackage{wasysym}
\usepackage[T1]{fontenc}
\newcommand{\be}{\begin{equation}}
\newcommand{\ee}{\end{equation}}
\newcommand{\beq}{\begin{eqnarray}}
\newcommand{\eeq}{\end{eqnarray}}

\newcommand{\Ta}{T^{(a)}}
\newcommand{\bTa}{\bar{T}^{(a)}}
\newcommand{\bTb}{\bar{T}^{(b)}}
\newcommand{\bTc}{\bar{T}^{(c)}}

\newcommand{\Tua}{T^{(1,a)}}
\newcommand{\Tdb}{T^{(2,b)}}
\newcommand{\Ttc}{T^{(3,c)}}
\newcommand{\Tl}{\bar{T}^{L(ab)}}
\newcommand{\Tr}{\bar{T}^{R(ab)}}
\newcommand{\Tlac}{\bar{T}^{L(ac)}}
\newcommand{\Tlbc}{\bar{T}^{L(bc)}}
\newcommand{\Trac}{\bar{T}^{R(ac)}}
\newcommand{\Trbc}{\bar{T}^{R(bc)}}

\newcommand{\qq}{\quad\quad}

\begin{document}

\title{Non-universal relativistic kinematics}

\author{J.M. Carmona}
\affiliation{Departamento de F\'{\i}sica Te\'orica,
Universidad de Zaragoza, Zaragoza 50009, Spain}
\author{J.L. Cort\'es}
\affiliation{Departamento de F\'{\i}sica Te\'orica,
Universidad de Zaragoza, Zaragoza 50009, Spain}
\author{B. Romeo}
\email{jcarmona@unizar.es, cortes@unizar.es, bromeo@unizar.es}
\affiliation{Departamento de F\'{\i}sica Te\'orica,
Universidad de Zaragoza, Zaragoza 50009, Spain}

\begin{abstract}
We present a systematic derivation of the constraints that the relativity principle imposes
between coefficients of a deformed (but rotational invariant) momentum composition law, dispersion relation, and
momentum transformation laws, at first order in a power expansion of an ultraviolet energy scale. This work
generalizes previous results to the case of particle-dependent relativistic kinematics, which can have interesting
phenomenological applications that we explore in the second part of the manuscript.
\end{abstract}

\maketitle

\section{Introduction}

The relativity principle (RP), or the equivalence between a class of observers (the inertial frames) related by a (10-parameter) set of transformations, has been at the core of all physical theories describing Nature up to date. In fact it was the RP, taken as a fundamental ingredient of classical mechanics, together with the validity of the laws of electrodynamics, what guided Einstein to propose special relativity (SR). At present, the symmetry of SR, Lorentz invariance (LI), is a basic constituent of relativistic quantum field theories, which account with great success for the laws of elementary particle physics.

However, some approaches to quantum gravity have suggested hints of Lorentz invariance violation (LIV)~\cite{Kostelecky1989,*Amelino-Camelia1998,*Gambini1999,*Carroll2001,*Lukierski1995,*AmelinoCamelia:1999pm,*Burgess:2002tb,*Barcelo:2005fc}, which could lead one to think about the necessity to abandon the RP. Fortunately, this is not necessarily the case: it is possible to go beyond SR and still maintain a relativistic theory. A specific realization of this idea is given by doubly special relativity (DSR) theories~\cite{Amelino-Camelia2001,*AmelinoCamelia:2001vy,*KowalskiGlikman:2000dz,%
*Amelino-Camelia2002a,*Amelino-Camelia2002b,*Magueijo:2002am,*Judes:2002bw,*Magueijo:2002xx,*Heyman:2003hs,*AmelinoCamelia:2010pd}. DSR considers deformed Lorentz transformations between inertial frames which preserve the form of a modified dispersion relation (MDR),
in which a high-energy scale $\Lambda$ (usually, the Planck mass) appears. The presence of this energy scale requires (by simple dimensional arguments) that the deformed transformations act nonlinearly in momentum space. This has another consequence: for systems of more than one particle, the usual linear energy-momentum conservation law is not compatible with the RP (that is, with the nonlinear Lorentz transformations), so that in order to give relativistic conservation laws one has to define a modified composition law (MCL) beyond the simple addition of the energies and momenta of the particles in the system.

The general conclusion is that in a relativistic theory beyond SR, the RP imposes restrictions between the deformed Lorentz transformations, the MDR, and the MCL. This was explicitely worked out in several examples in Ref.~\cite{AmelinoCamelia:2011yi}, where a requirement necessary for ``DSR compatibility'' of nonlinear composition laws on momentum-space was identified, as a ``golden-rule'' between the coefficients of a deformed dispersion relation and the composition laws when working at leading order in the scale of the deformation (the inverse of $\Lambda$). This relation was seen to be a necessary, but not sufficient, condition in Ref.~\cite{Carmona2012}, where a generalization of the examples presented in Ref.~\cite{AmelinoCamelia:2011yi} was derived. Both Refs.~\cite{AmelinoCamelia:2011yi} and~\cite{Carmona2012} assumed a universal deformation for all particles.

Non-universal relativistic kinematics was considered for the first time in Ref.~\cite{Amelino-Camelia2012}, presenting once more specific examples of compatibility between deformed Lorentz transformations, dispersion relations and composition laws. The kinematics of SR is of course universal, but there are several motivations to consider non-universality in a generalization of SR: firstly, as a quantum-gravity effect, not all systems have to be necessarily affected in the same way by the quantum spacetime structure (in fact, one would expect that the coefficients characterizing the deformation were renormalized differently even if the fundamental Lorentz violation is universal~\cite{Alfaro:2004aa,*Mattingly:2005re}); secondly, it might be relevant in the description of composite particles, such as atoms or, in general, macroscopic bodies (and offer therefore a solution to the soccer-ball or the spectator problems which are commonly present in generalized relativistic kinematics, see e.g.~\cite{AmelinoCamelia:2005xr}). A final motivation would be the possible phenomenological interest of this scenario: since the sensitivity of our observations to departures from SR kinematics differs by orders of magnitude when one considers different systems, in the case of a non-universal kinematics the stringent limits obtained, for example, on a possible energy dependence of the velocity of propagation of photons, or on the difference between the velocity of propagation of ultrarelativistic electrons and photons, would not necessarily apply to other particles like neutrinos or the dark matter sector, where one could have much larger departures from SR kinematics without entering into conflict with our limited observations in these systems.

The aim of the present work is to generalize the findings of Ref.~\cite{Amelino-Camelia2012} in a similar way as Ref.~\cite{Carmona2012} represented a systematic derivation of the results of Ref.~\cite{AmelinoCamelia:2011yi} for the universal case. The examples of non-universal kinematics of Ref.~\cite{Amelino-Camelia2012} should then correspond to particular choices of coefficients of the general framework presented here.

We will present the generic construction for a non-universal kinematics beyond SR in Sec. II, where we also derive a generalization of the ``golden-rules'' previously found for the universal case, and will apply them to the simple case in which the non-universality is reduced to the existence of two types of particles: this is what we call a bipartite relativistic kinematics (BRK). As we will see, some examples appeared previously in the literature are specific realizations of a BRK. Sec. III will be devoted to applications of a non-universal relativistic kinematics to different physical processes. We will consider the case of thresholds in two-particle decays and the ultrarelativistic limit of two-body scattering processes and see how the consequences of the presence of a relativity principle in a kinematics beyond special relativity are physically distinguishible from a Lorentz violating scenario. A detailed analysis will be done in the simple case of an elastic scattering between two particles in a BRK scenario which, as we will see, might be phenomenologically relevant in a context in which the modification in the kinematics only affect (or is more relevant) to elementary particles, while the corrections are smaller for composite objects. A discussion and some comments will then be given in Sec. IV.

\section{Non-universal relativistic kinematics}

\subsection{General discussion}

We start by considering a generalized relativistic kinematics in the one-particle sector (what we may denote as RP$_1$, RP standing for Relativity Principle). The momentum of a particle of type $(a)$ satisfies a modified dispersion relation
\begin{equation}
C^{(a)}(p)=p_0^2-\vec{p}^2+\frac{\alpha_1^a}{\Lambda}p_0^3+\frac{\alpha_2^a}{\Lambda}p_0\vec{p}^2=m^2\,.
\label{eq:MDR}
\end{equation}
This is the most general expression ($\alpha_1^a$ and $\alpha_2^a$ are dimensionless coefficients) which is a polynomial in the components of the four-momentum $(p_0,p_i)$, satisfies rotational invariance, and extends the special-relativistic expression $p_0^2-\vec{p}^2=m^2$ to modifications of order $1/\Lambda$.

The momentum $p$ transforms under a boost by means of a deformed Lorentz transformation
\begin{equation}
	p\to \Ta(p)=T(p)+\bTa(p),
\label{eq:T-one}
\end{equation}
where $T(p)$ is the usual Lorentz boost, which we write infinitesimally as
\begin{equation}
	[T(p)]_0=p_0+ \vec{p}\cdot\vec{\xi} \quad \quad \quad \quad [T(p)]_i=p_i+p_0 \,\xi_i\,,
\label{eq:T}
\end{equation}
where $\vec{\xi}$ is the vector parameter of the boost. The most general expression for $\bTa(p)$ turns out to be (see Ref.~\cite{Carmona2012})\footnote{As it is derived in Ref.~\cite{Carmona2012}, the coefficients of the different terms in Eq.~(\ref{eq:barT}) are obtained after one imposes the condition that the modified boosts reproduce the Lorentz algebra, ie, that the commmutator of two boosts corresponds to a rotation.}:
\begin{equation}
	[\bTa(p)]_0=\frac{\lambda_1^a}{\Lambda}p_0\,(\vec{p}\cdot\vec{\xi}) \quad\quad \quad \quad [\bTa(p)]_i=\frac{\lambda_2^a}{\Lambda}p_0^2\,\xi_i+\frac{\lambda_3^a}{\Lambda}	\vec{p}^2\,\xi_i+\frac{\lambda_1^a+2\lambda_2^a+2\lambda_3^a}{\Lambda}p_i \,(\vec{p}\cdot\vec{\xi})\,.
\label{eq:barT}
\end{equation}
where again the $\lambda_i^a$ are dimensionless coefficients.

When one imposes invariance of the MDR under the generalized boosts, $C^{(a)}(p)=C^{(a)}\left(\Ta(p)\right)$, then the following relation between the coefficients of the MDR and the generalized boost is obtained:
\begin{equation}
\alpha^a_1 = -2 \,(\lambda^a_1 + \lambda^a_2 + 2\lambda^a_3) \quad \quad \quad \quad
\alpha^a_2 = 2 \, (\lambda^a_1 + 2\lambda^a_2 + 3\lambda^a_3)\,.
\label{eq:alpha(lambda)}
\end{equation}

Let us now consider the two-particle system formed by a particle of type $(a)$ and a particle of type $(b)$ (we denote this sector as RP$_2$), of momenta $p$ and $q$, respectively. The general form of a composition law compatible with rotational invariance is
\begin{equation}
\left[p\oplus q\right]_0 = p_0 + q_0 + \frac{\beta^{ab}_1}{\Lambda} \, p_0 q_0 + \frac{\beta^{ab}_2}{\Lambda} \, \vec{p}\cdot\vec{q} \quad \quad \quad \quad \left[p \oplus q\right]_i = p_i + q_i + \frac{\gamma^{ab}_1}{\Lambda} \, p_0 q_i + \frac{\gamma^{ab}_2}{\Lambda} \, p_i q_0
+ \frac{\gamma^{ab}_3}{\Lambda} \, \epsilon_{ijk} p_j q_k
\label{eq:MCL}
\end{equation}
where $\epsilon_{ijk}$ is the Levi-Civita symbol, a totally antisymmetric tensor, and it is implemented the condition
\begin{equation}
p \oplus q|_{q=0} = p \quad \quad  \quad \quad p \oplus q|_{p=0} = q\,.
\label{eq:cl0}
\end{equation}
Observe that the MCL mixes components of $p$ and $q$ in its terms; therefore, in order to have generalized boosts compatible with it, they will have to depend on both momenta. Also, the order of the momenta in the MCL Eq.~\eqref{eq:MCL} is relevant, so that the transformations on $p$ and $q$ will in general depend on that order. We define then a boost on RP$_2$ as
\begin{equation}
\{p,q\} \to \{\Tua(p,q),\Tdb(p,q)\}\,,
\label{eq:rp2transf}
\end{equation}
where the superindex $(1,a)$ in $\Tua$ indicates that it corresponds to the transformation on the first momentum of the ordered set $\{p,q\}$, which corresponds to a particle of type $(a)$, and we explicitly write that the transformation depends on both momenta $p$ and $q$. The expressions of $\Tua(p,q)$ and $\Tdb(p,q)$ are
\begin{equation}
\Tua(p,q)=T(p)+\bTa(p)+\Tl(p,q) \quad \quad \quad \quad \Tdb(p,q)=T(q)+\bTb(q)+\Tr(p,q)\,,
\label{eq:T-two}
\end{equation}
where $T$ and $\bar T$ were defined in Eqs.~(\ref{eq:T}) and~(\ref{eq:barT}), respectively, and $\Tl$, $\Tr$ have the general expressions\footnote{In order to obtain them, one has to impose once more the invariance of the dispersion relation, $C^{(a)}(p)=C^{(a)}(\Tua(p,q))$, $C^{(b)}(q)=C^{(b)}(\Tdb(p,q))$, and the consistency with the Lorentz algebra, see Ref.~\cite{Carmona2012}.}:
\begin{align}
\left[\Tl(p,q)\right]_0&=\frac{\eta_1^{Lab}}{\Lambda}\,q_0\,(\vec{p}\cdot\vec{\xi})+\frac{\eta_2^{Lab}}{\Lambda}\,(\vec{p}\wedge\vec{q})\cdot\vec{\xi} \quad \quad \quad \quad 
\left[\Tr(p,q)\right]_0=\frac{\eta_1^{Rab}}{\Lambda}\,p_0\,(\vec{q}\cdot\vec{\xi})+\frac{\eta_2^{Rab}}{\Lambda}\,(\vec{q}\wedge\vec{p})\cdot\vec{\xi} \nonumber \\
\left[\Tl(p,q)\right]_i&=\frac{\eta_1^{Lab}}{\Lambda}\,q_0 p_0 \xi_i + \frac{\eta_2^{Lab}}{\Lambda}\,(p_0\epsilon_{ijk}q_j\xi_k-q_0\epsilon_{ijk}p_j\xi_k)+\frac{\eta_1^{Lab}}{\Lambda}(q_i\,\vec{p}\cdot\vec{\xi}-\xi_i\,\vec{q}\cdot \vec{p}) \nonumber \\
\left[\Tr(p,q)\right]_i&=\frac{\eta_1^{Rab}}{\Lambda}\,p_0 q_0 \xi_i + \frac{\eta_2^{Rab}}{\Lambda}\,(q_0\epsilon_{ijk}p_j\xi_k-p_0\epsilon_{ijk}q_j\xi_k)+\frac{\eta_1^{Rab}}{\Lambda}(p_i\,\vec{q}\cdot\vec{\xi}-\xi_i\,\vec{p}\cdot \vec{q}).
\label{eq:barTLR}
\end{align}

The principle of relativity now establishes a relationship between the coefficients of the MCL ($\beta^{ab}_1$,$\beta^{ab}_2$,$\gamma^{ab}_1$,$\gamma^{ab}_2$,$\gamma^{ab}_3$) and the coefficients of the generalized boosts ($\lambda^{ab}_1$, $\lambda^{ab}_2$, $\lambda^{ab}_3$, $\eta_1^{Lab}$, $\eta_1^{Rab}$, $\eta_2^{Lab}$, $\eta_2^{Rab}$). In order to get this relation, we follow Ref.~\cite{Amelino-Camelia2012} and consider the covariance of the conservation law:
\begin{equation}
p\oplus q =0 \quad \Rightarrow \quad \Tua(p,q)\oplus \Tdb(p,q)=0\,.
\label{eq:CL}
\end{equation}

The condition $p\oplus q=0$ allows one to express $q$ in terms of $p$:
\begin{equation}
q_0=-p_0+\frac{\beta_1^{ab}}{\Lambda} p_0^2+\frac{\beta_2^{ab}}{\Lambda}\vec{p}^2\, \quad \quad \quad \quad q_i=-p_i+\frac{\gamma_1^{ab}+\gamma_2^{ab}}{\Lambda}p_0 p_i\,.
\label{eq:qandp}
\end{equation}
From Eq.~(\ref{eq:T-two}), $\Tua(p,q)\oplus \Tdb(p,q)=0$ implies that (note that the composition $\oplus$ is the ordinary sum if one of the addends is of order $1/\Lambda$):
\begin{equation}
T(p)\oplus T(q)=-[\bTa(p)+\bTb(q)+\Tl(p,q)+\Tr(p,q)]\,;
\label{eq:oplustoplus}
\end{equation}
however, since $\bar{T}$ starts at order $1/\Lambda$, we can use the relation between $q$ and $p$ at zero order, that is, $q=-p+\mathcal{O}(\Lambda^{-1})$, so that
\begin{align}
&\bTb(q)=\bTb(-p)=\bTb(p) \nonumber \\ 
&\Tl(p,q)=\Tl(p,-p)=-\Tl(p,p) \quad \quad \quad \Tr(p,q)=\Tr(p,-p)=-\Tr(p,p)\,.
\end{align}
We end up with the following relation:
\begin{equation}
T(p)\oplus T(q)=\Tl(p,p)+\Tr(p,p)-\bTa(p)-\bTb\,(p).
\label{eq:TpandTq}
\end{equation}
The temporal and spacial parts of the left hand side of the previous equation can be expanded according to Eq.~\eqref{eq:MCL}, using also Eq.~\eqref{eq:T} and the relation between $q$ and $p$ to order $1/\Lambda$ [Eq.~\eqref{eq:qandp}] in those terms which contain $T(q)$ at order zero in the $1/\Lambda$ expansion. Also, the right hand side of Eq.~\eqref{eq:TpandTq} can be elaborated using Eqs.~\eqref{eq:barT} and~\eqref{eq:barTLR}, so that comparing both sides of Eq.~\eqref{eq:TpandTq} one arrives to three independent conditions:
\begin{align}
\lambda_1^a+\lambda_1^b-(\eta_1^{Lab}+\eta_1^{Rab})& =-(\gamma_1^{ab}+\gamma_2^{ab})+2(\beta_1^{ab}+\beta_2^{ab}) \nonumber \\
\lambda_2^a+\lambda_2^b-(\eta_1^{Lab}+\eta_1^{Rab})& =(\gamma_1^{ab}+\gamma_2^{ab})-\beta_1^{ab} \nonumber \\
\lambda_3^a+\lambda_3^b+(\eta_1^{Lab}+\eta_1^{Rab})& =-\beta_2^{ab} \,.
\label{eq:lambdaconditions}
\end{align}
If we particularize the previous relations to the case in which both momenta are of type $(a)$, we get
\begin{equation}
\lambda_1^a-\frac{\eta_1^{La}+\eta_1^{Ra}}{2}=-\frac{\gamma_1^a+\gamma_2^a}{2}+(\beta_1^a+\beta_2^a) \quad  \quad \quad  \quad
\lambda_2^a-\frac{\eta_1^{La}+\eta_1^{Ra}}{2}=\frac{\gamma_1^a+\gamma_2^a}{2}-\frac{\beta_1^a}{2} \quad  \quad \quad  \quad
\lambda_3^a+\frac{\eta_1^{La}+\eta_1^{Ra}}{2}=-\frac{\beta_2^a}{2}\,,
\label{eq:relsRP2a}
\end{equation}
where we have introduced the notation $\eta_1^{Laa}=\eta_1^{La}$, $\eta_1^{Raa}=\eta_1^{Ra}$, $\beta_i^{aa}=\beta_i^a$, $\gamma_i^{aa}=\gamma_i^a$, and there are analogous relations by exchanging the labels $a$ and $b$. 

Combining Eqs.~(\ref{eq:alpha(lambda)}) and~\eqref{eq:relsRP2a}, one gets the ``golden rules''
\begin{equation}
\alpha^a_1 = - \beta^a_1 \quad \quad \quad \quad \alpha^a_2 = \gamma^a_1 + \gamma^a_2 - \beta^a_2 \,,
\label{eq:golden}
\end{equation}
which are the same relations as the ones obtained in Ref.~\cite{Carmona2012} for the case of universal kinematics.

By replacing expressions~\eqref{eq:relsRP2a} for $\lambda_i^a$ and the equivalent ones for $\lambda_i^b$ into Eq.~(\ref{eq:lambdaconditions}), it is obtained
\begin{align}
\begin{split}
\frac{(\eta_1^{La}+\eta_1^{Ra})}{2}+\frac{(\eta_1^{Lb}+\eta_1^{Rb})}{2}-(\eta_1^{Lab}+\eta_1^{Rab})=&-\left[\left(\gamma_1^{ab}-\frac{\gamma_1^a+\gamma_1^b}{2}\right)+\left(\gamma_2^{ab}-\frac{\gamma_2^a+\gamma_2^b}{2}\right)\right] \\ 
&+ 2 \left[\left(\beta_1^{ab}-\frac{\beta_1^a+\beta_1^b}{2}\right)+\left(\beta_2^{ab}-\frac{\beta_2^a+\beta_2^b}{2}\right)\right]
\end{split}
\nonumber \\
\frac{(\eta_1^{La}+\eta_1^{Ra})}{2}+\frac{(\eta_1^{Lb}+\eta_1^{Rb})}{2}-(\eta_1^{Lab}+\eta_1^{Rab})=&
\left[\left(\gamma_1^{ab}-\frac{\gamma_1^a+\gamma_1^b}{2}\right)+\left(\gamma_2^{ab}-\frac{\gamma_2^a+\gamma_2^b}{2}\right)\right] - \left(\beta_1^{ab}-\frac{\beta_1^a+\beta_1^b}{2}\right)
\nonumber \\
-\frac{(\eta_1^{La}+\eta_1^{Ra})}{2}-\frac{(\eta_1^{Lb}+\eta_1^{Rb})}{2}+(\eta_1^{Lab}+\eta_1^{Rab})=&
- \left(\beta_2^{ab}-\frac{\beta_2^a+\beta_2^b}{2}\right)\,.
\end{align}

From the last equations, one works out two expressions for $\beta_i^{ab}$ and $\gamma_i^{ab}$ in terms of $\beta_i^a$, $\beta_i^b$, $\gamma_i^a$ and $\gamma_i^b$:
\begin{align}
\beta_1^{ab}-\frac{\beta_1^a+\beta_1^b}{2}&=0 \label{eq:betaab}\\
\left[\left(\gamma_1^{ab}-\frac{\gamma_1^a+\gamma_1^b}{2}\right)+\left(\gamma_2^{ab}-\frac{\gamma_2^a+\gamma_2^b}{2}\right)\right] - \left(\beta_2^{ab}-\frac{\beta_2^a+\beta_2^b}{2}\right)&=0
\label{eq:gammaab}
\end{align}
and an equation for $\eta_1^{Lab}+\eta_1^{Rab}$:
\begin{equation}
\eta_1^{Lab}+\eta_1^{Rab}-\frac{(\eta_1^{La}+\eta_1^{Ra})}{2}-\frac{(\eta_1^{Lb}+\eta_1^{Rb})}{2}=-\left(\gamma_1^{ab}-\frac{\gamma_1^a+\gamma_1^b}{2}\right)-\left(\gamma_2^{ab}-\frac{\gamma_2^a+\gamma_2^b}{2}\right)=-\left(\beta_2^{ab}-\frac{\beta_2^a+\beta_2^b}{2}\right)\,.
\label{eq:etaunoab}
\end{equation}

Eqs.~\eqref{eq:lambdaconditions} are not, however, the full set of relations between the coefficients of the MCL and the generalized boosts required by the RP. To get all of them, it is necessary to consider a three-particle system, RP$_3$, in which boosts act
\begin{equation}
\{p,q,k\} \to \{\Tua(p,q,k),\Tdb(p,q,k),\Ttc(p,q,k)\}\,,
\end{equation}
where
\begin{align}
\Tua(p,q,k)&=T(p)+\bTa(p)+\Tl(p,q)+\Tlac(p,k) \\	
\Tdb(p,q,k)&=T(q)+\bTb(q)+\Tr(p,q)+\Tlbc(q,k) \\
\Ttc(p,q,k)&=T(k)+\bTc(k)+\Trac(p,k)+\Trbc(q,k)\,.
\end{align}
At order $1/\Lambda$, all nonlinear terms need to be quadratic in momenta, so that we see that for a system of an arbitrary number of particles, there are no new ingredients in the generalized Lorentz transformations from those appearing in the two-particle system. However, the need to go to RP$_3$ is natural, since the conservation law in RP$_2$ $p\oplus q=0$ implies that only one momentum is independent. Since the $1/\Lambda$ terms are quadratic in momenta, one needs two independent momenta to get all the constraints between coefficients of the MCL and the generalized boosts.

The composition of the three momenta $p$, $q$ and $k$, of type $(a)$, $(b)$ and $(c)$, respectively, is completely determined by the coefficients of the composition of two momenta, since the sum $p\oplus q\oplus k$ has to reduce to the composition of two momenta when the other one is equal to zero. We have then:
\begin{align}
[p \oplus q \oplus k]_0 &= p_0 + q_0 + k_0 + \frac{\beta_1^{ab}}{\Lambda}p_0 q_0+ \frac{\beta_1^{ac}}{\Lambda}p_0 k_0 + \frac{\beta_1^{bc}}{\Lambda}q_0 k_0+
\frac{\beta_2^{ab}}{\Lambda}\vec{p}\cdot \vec{q} + \frac{\beta_2^{ac}}{\Lambda}\vec{p}\cdot \vec{k} + \frac{\beta_2^{bc}}{\Lambda}\vec{q}\cdot \vec{k} \\
[p \oplus q \oplus k]_i &= p_i + q_i + k_i + \frac{\gamma_1^{ab}}{\Lambda}p_0 q_i+ \frac{\gamma_1^{ac}}{\Lambda}p_0 k_i + \frac{\gamma_1^{bc}}{\Lambda}q_0 k_i+
\frac{\gamma_2^{ab}}{\Lambda}p_i q_0+ \frac{\gamma_2^{ac}}{\Lambda}p_i k_0 + \frac{\gamma_2^{bc}}{\Lambda}q_i k_0 \nonumber \\
& +  
\frac{\gamma_3^{ab}}{\Lambda} \epsilon_{ijl} p_j q_l + \frac{\gamma_3^{ac}}{\Lambda} \epsilon_{ijl} p_j k_l  + \frac{\gamma_3^{bc}}{\Lambda} \epsilon_{ijl} q_j k_l 
\label{eq:3MCL}
\end{align}

Proceeding as in RP$_2$, $\Tua(p,q,k)\oplus \Tdb(p,q,k)\oplus \Ttc(p,q,k)=0$, for momenta $p,q,k$ verifying $p\oplus q\oplus k=0$, implies that
\begin{equation}
\begin{split}
T(p)\oplus T(q)\oplus T(k)&=\Tl(p,p)+\Tr(p,p)+\Tlbc(k,k)+\Trbc(k,k)\\ &+\Tl(p,k)+\Tlbc(p,k)-\Tlac(p,k) +\Tr(p,k)\\&+\Trbc(p,k)-\Trac(p,k)-\bTa(p)-\bTc(k)-\bTb(p+k)\,.
\label{eq:condRP3-NU}
\end{split}
\end{equation}
Expanding both sides of this equality, and after some algebra, one arrives to a set of eleven equations: the three relations obtained in the context of RP$_2$, Eq.~(\ref{eq:lambdaconditions}), for every pair of particles $(ab)$, $(ac)$ or $(bc)$, 
%which could be written as three equations for $\beta_i^{ab}$, $\gamma_i^{ab}$, and $\eta_1^{Lab}+\eta_1^{Rab}$, Eqs.~(\ref{eq:betaab}), (\ref{eq:gammaab}), and~\eqref{eq:etaunoab}, 
plus two new conditions
\begin{align}
(\eta_1^{Lab}-\eta_1^{Rab})+(\eta_1^{Lbc}-\eta_1^{Rbc})-(\eta_1^{Lac}-\eta_1^{Rac})&=(\gamma_2^{ab}-\gamma_1^{ab})+(\gamma_2^{bc}-\gamma_1^{bc})-(\gamma_2^{ac}-\gamma_1^{ac}) 
\label{eq:eta-NU1} \\
(\eta_2^{Lab}-\eta_2^{Rab})+(\eta_2^{Lbc}-\eta_2^{Rbc})-(\eta_2^{Lac}-\eta_2^{Rac})&=\gamma_3^{ab}+\gamma_3^{bc}-\gamma_3^{ac}\,.
\label{eq:eta-NU2}
\end{align}

In summary, Eqs.~(\ref{eq:betaab}) and (\ref{eq:gammaab}) for every pair of particle types are the consistency conditions for the parameters appearing in the composition laws between particles of different types, $\beta^{ab}$ and $\gamma^{ab}$, and Eqs.~\eqref{eq:etaunoab}, \eqref{eq:eta-NU1} and~\eqref{eq:eta-NU2} are the conditions for the parameters $\eta_1^L,\eta_1^R,\eta_2^L,\eta_2^R$ appearing in the transformations laws. 

Besides this, Eq.~\eqref{eq:relsRP2a}, together with Eqs.~\eqref{eq:eta-NU1} and \eqref{eq:eta-NU2}, particularized for the specific case in which all three particles are of the same type, allow one to write the coefficients of the MCL in terms of the parameters of the generalized Lorentz boosts in the sector of particles of a single type,
\begin{alignat}{3}
\beta_1^a &= 2 \,(\lambda_1^a + \lambda_2^a + 2\lambda_3^a) \quad\quad &
\beta_2^a &= -2 \lambda_3^a - \eta_1^{La} - \eta_1^{Ra} \quad\quad & \nonumber \\
\gamma_1^a &= \lambda_1^a + 2\lambda_2^a + 2\lambda_3^a - \eta_1^{La} \quad\quad &
\gamma_2^a &= \lambda_1^a + 2\lambda_2^a + 2\lambda_3^a - \eta_1^{Ra} \quad\quad & \gamma_3^a = \eta_2^{La} - \eta_2^{Ra} \,.
\label{MCL-GLT}
\end{alignat}
Alternatively, there exists a biparametric family of implementations of Lorentz transformations in RP$_1$ (all of them equivalent) that guarantee a compatibility with a given MCL and the RP. This of course reproduces the situation present in the universal kinematics of Ref.~\cite{Carmona2012}.

In the next Subsection we will consider the simplest example of non-universality, containing particles of only two types.

\subsection{A particular case: bipartite relativistic kinematics (BRK)}
\label{sec-BRK}

Let us rewrite the conditions that the relativistic principle imposes for the paremeters of the MCL and the modified transformation laws in the case of only two types of particles (bipartite relativistic kinematics, or BRK), types $(a)$ and $(b)$.

The MCL of momenta of type $(a)$ [or of type $(b)$] is given by the coefficients $\beta_i^a$ and $\gamma_i^a$ (or $\beta_i^b$ and $\gamma_i^b$). For the composition of a momentum of type $(a)$ and another of type $(b)$, we will need the coefficients $\beta_i^{ab}$ and $\gamma_i^{ab}$; however, while $\gamma_1^{ab}$ and $\gamma_2^{ab}$ are free parameters, $\beta_1^{ab}$ and $\beta_2^{ab}$ are completely determined by [see Eqs.\eqref{eq:betaab} and~\eqref{eq:gammaab}]:
\begin{align}
\beta_1^{ab}&=\frac{\beta_1^a+\beta_1^b}{2} \label{beta1ab} \\
\beta_2^{ab}&=\frac{\beta_2^a+\beta_2^b}{2}+\left(\gamma_1^{ab}-\frac{\gamma_1^a+\gamma_1^b}{2}\right)+\left(\gamma_2^{ab}-\frac{\gamma_2^a+\gamma_2^b}{2}\right). \label{beta2ab}
\end{align}

For the parameters of the transformation law of a system of particles of different type, $\eta_1^{Lab}$, $\eta_1^{Rab}$, $\eta_2^{Lab}$, $\eta_2^{Rab}$, we have the following conditions [see Eqs.~\eqref{eq:etaunoab}, \eqref{eq:eta-NU1} and~\eqref{eq:eta-NU2}]:
\begin{align}
\eta_1^{Lab}+\eta_1^{Rab}&=\frac{\eta_1^{La}+\eta_1^{Lb}}{2}+\frac{\eta_1^{Ra}+\eta_1^{Rb}}{2}-\left(\beta_2^{ab}-\frac{\beta_2^a+\beta_2^b}{2}\right) \label{etasimab}\\
(\eta_1^{Lab}-\eta_1^{Rab})+(\eta_1^{Lba}-\eta_1^{Rba})&=(\gamma_2^{ab}-\gamma_1^{ab})+(\gamma_2^{ba}-\gamma_1^{ba}) \label{etaantisim1}\\
(\eta_2^{Lab}-\eta_2^{Rab})+(\eta_2^{Lba}-\eta_2^{Rba})&=\gamma_3^{ab}+\gamma_3^{ba}\,. \label{etaantisim2}
\end{align}

Eqs.~\eqref{etaantisim1} and~\eqref{etaantisim2} were obtained form Eqs.~\eqref{eq:eta-NU1} and~\eqref{eq:eta-NU2} by taking $a=c$ and making use of Eq.~\eqref{MCL-GLT}. Note that if one takes $a=b$ (or $b=c$) in Eqs.~\eqref{etaantisim1} and~\eqref{etaantisim2}, then one just reproduces relations already contained in Eq.~\eqref{MCL-GLT}.

Eq.~\eqref{beta1ab} tells us that $\beta_1^{ba}=\beta_1^{ab}$, but this is not necessarily the case for the rest of coefficients. We get for $\beta_2^{ba}$ and $(\eta_1^{Lba}+\eta_1^{Rba})$ similar equations to Eqs.~\eqref{beta2ab} and~\eqref{etasimab}, while Eqs.~\eqref{etaantisim1} and~\eqref{etaantisim2} are symmetric under the exchange of the $a$ and $b$ labels. On the other hand, the $\gamma_i^{ba}$ are free parameters, as it was the case for the $\gamma_i^{ab}$.

A particularly simple choice of BRK is one in which the relation \eqref{beta1ab} is extended for all the remaining coefficients of the generalized kinematics; in particular, this makes that the $(ba)$ coefficients are equal to the $(ab)$ ones. It is notable that all the conditions we got are compatible with this choice. In this case, one would only need to specify the generalized kinematics for both sectors separately, and then this would determine the kinematics when particles of both types are present. As we will see in the next Section, the examples discussed in Ref.~\cite{Amelino-Camelia2012} correspond to this specific choice. 
%\begin{alignat}{2}
%\beta_1^{ab}&=\frac{\beta_1^a+\beta_1^b}{2} \quad \quad \quad \beta_2^{ab}&=\frac{\beta_2^a+\beta_2^b}{2} & \quad \quad \quad \\
%\gamma_1^{ab}&=\frac{\gamma_1^a+\gamma_1^b}{2} \quad \quad \quad  \gamma_2^{ab}&=\frac{\gamma_2^a+\gamma_2^b}{2} \quad \quad \quad  \gamma_3^{ab}&=\frac{\gamma_3^a+\gamma_3^b}{2} \\
%\eta_1^{Lab}&=\frac{\eta_1^{La}+\eta_1^{Lb}}{2} \quad \quad \quad \eta_1^{Rab}&=\frac{\eta_1^{Ra}+\eta_1^{Rb}}{2} \quad \quad \quad
%\eta_2^{Lab}&=\frac{\eta_2^{La}+\eta_2^{Lb}}{2} \quad \quad \quad \eta_2^{Rab}=\frac{\eta_2^{Ra}+\eta_2^{Rb}}{2}
%\end{alignat}

\subsection{Previous examples of non-universal kinematics}

As we mentioned in the Introduction, Reference~\cite{Amelino-Camelia2012} was the first to introduce a non-universal kinematics in the presence of a relativity principle. It did so by exploring some specific examples, but it missed a systematic derivation of the relations between coefficients of the modified dispersion relations and the different composition laws as the one presented here. We will now show that the examples presented there are indeed particular cases of our general discussion (in fact all of them contain only two sectors of particles, and are therefore examples of what we have called bipartite relativistic kinematics in the previous subsection).

\subsubsection{The simplest case with commutative composition of momenta}

The first example presented in Ref.~\cite{Amelino-Camelia2012} contains two sectors of particles: momenta of the first kind $(a)$ satisfy
\begin{equation}
m^2 = p_0^2 - p_j^2 + 2\ell p_0 p_j^2~,
\label{metricHLp}
\end{equation}
where the deformation scale $\ell$ plays the role of $1/\Lambda$, and the composition laws are
\begin{equation}
(p \oplus_{\ell} p^\prime)_j = p_j + p^\prime_j + \ell p_0 p^\prime_j  + \ell p^\prime_0 p_j ~,~~~
(p \oplus_{\ell} p^\prime)_0 = p_0 + p^\prime_0~,
\label{connectionHLp}
\end{equation}
while momenta of the second kind $(b)$ (denoted by $k$) satisfy the dispersion relation and composition laws of special relativity. 
This is a very simple case for which one sector is trivial and the other has a commutative composition of momenta.
Then (through a ``trial and error exercise'') it is presented a possible ``mixing composition law'' that is consistent with the relativity principle:
\begin{equation}
(p \oplus k)_j = p_j + k_j + \frac{\ell}{2} p_0 k_j + \frac{\ell}{2} k_0 p_j ~,~~~
(p \oplus k)_0 = p_0 + k_0~.
\label{connectionHLpk}
\end{equation}
In our notation, therefore:
\begin{align}
\alpha_2^a& =2 \qq \alpha_1^a=\beta_1^a=\beta_2^a=\gamma_3^a=0 \qq \gamma_1^a=\gamma_2^a=1 \\
\alpha_1^b& = \alpha_2^b = \beta_1^b = \beta_2^b = \gamma_1^b=\gamma_2^b=\gamma_3^b = 0 \\
\beta_1^{ab}& =\beta_2^{ab}=0 \qq \gamma_1^{ab}=\gamma_2^{ab}=\frac{1}{2} \qq \gamma_3^{ab}=0\,.
\label{eq:ex1}
\end{align}
It is then immediate to check that both particle sectors satisfy individually the golden rules Eq.~\eqref{eq:golden}. Eq.~\eqref{beta1ab} is trivially satisfied since the composition of energies inside and between both sectors is that of SR, and Eq.~\eqref{beta2ab} is also satisfied with the particular choice $\gamma_1^{ab}=(\gamma_1^a+\gamma_1^b)/2$, and $\gamma_2^{ab}=(\gamma_2^a+\gamma_2^b)/2$.
We note also that Ref.~\cite{Amelino-Camelia2012} did not make any distinction between the composition $(p\oplus k)_j$ and $(k\oplus p)_j$, that is, it implicitly assumed that $\gamma_i^{ab}=\gamma_i^{ba}$ (which, as we remarked above, is an arbitrary choice). 

Ref.~\cite{Amelino-Camelia2012} also gives appropriate transformation laws for this example. For the first type (``$p$-type'') of particles the generators of boosts act as (considering the simpler $1+1$-dimensional case):
\begin{equation}
[N_{[p]}, p_0] =  p_1 - \ell p_0 p_1~,~~~
[N_{[p]}, p_1] =  p_0 + \ell p_0^2 + \ell p_1^2~,
\label{boostsHL}
\end{equation}
while for the second type (``$k$-type'') of particles, boosts act trivially: 
\begin{equation}
[N_{[k]}, k_0] =  k_1 ~,~~~
[N_{[k]}, k_1] =  k_0 ~.
\label{boostSRhl}
\end{equation}
In our language (see Eqs.~\eqref{eq:T} and~\eqref{eq:barT}):
\begin{equation}
\lambda_1^a=-1 \qq \lambda_2^a=1 \qq \lambda_3^a=0\,\qq \qq \lambda_1^b=\lambda_2^b=\lambda_3^b=0\,.
\label{eq:ex1lambda}
\end{equation}
Since the non-trivial composition of momenta of $p$-particles is commutative, Ref.~\cite{Amelino-Camelia2012} shows that the conservation law  $p \oplus_{\ell} p^\prime =  0$ has covariance ensured by the total-boost action
\begin{equation}
N_{[p \oplus_{\ell} p^\prime]} =N_{[p]}+N_{[p^\prime]}\, ,
\label{boostpp}
\end{equation}
which means that $\eta_1^{La}=\eta_1^{Ra}=\eta_2^{La}=\eta_2^{Ra}=0$. Indeed, Eqs.~\eqref{MCL-GLT} are satisfied for the particle sector $(a)$.

Finally, the previous work finds that also the ``mixed'' composition law $p\oplus k$ is compatible with a standard total-boost action:
\begin{equation}
N_{[p \oplus k]} =N_{[p]}+N_{[k]}~,
\label{boostpk}
\end{equation}
so that all the coefficients $\eta_i^{(L,R)ab}$ are zero and we see that Eqs.~\eqref{etasimab}, \eqref{etaantisim1} and \eqref{etaantisim2} are also satisfied.

Ref.~\cite{Amelino-Camelia2012} also checked that the ``tri-valent processes'' (following its language), which involve the composition of three momenta are covariant (the conservation laws are consistent with the RP). In fact, according to our analysis, this is automatic once relations Eq.~\eqref{eq:golden}, \eqref{beta1ab} and \eqref{beta2ab} are satisfied. 
 
\subsubsection{A ($\kappa$-Poincaré-inspired) more general scenario}

A less simple example is one in which none of the two sectors obeys the standard kinematics of SR and the composition of momenta is not commutative. Ref.~\cite{Amelino-Camelia2012} examines this situation in a $\kappa$-Poincaré-inspired scenario, in which 
both types of particles are governed by a $\kappa$-Poincaré-inspired DSR-deformation of Lorentz symmetry, but with different deformation scales $\ell$ and $\lambda$:\footnote{As a matter of fact Ref.~\cite{Amelino-Camelia2012} analyzes also a simpler example in which only one of the particle types has this $\kappa$-Poincaré-inspired DSR-deformation of Lorentz symmetry, while the other follows SR. In this scenario it explores a situation which is not included in the present study: the case in which $p\oplus k$ is not $k$ when $p\to 0$. This violates our Eq.~\eqref{eq:cl0} and, as Ref.~\cite{Amelino-Camelia2012} explains, is a feature that may be ``plausible but surprising''. Our general analysis excluded this situation for simplicity, but, in any case, as Ref.~\cite{Amelino-Camelia2012} also notes, this feature is not a general aspect of ``mixing composition laws'' that can be adopted in this particular example, and in fact one can adopt others inside the general framework considered here.}  
\begin{align}
m^2 = p_0^2 - p_j^2 + \ell p_0 p_j^2 & \qq \mu^2=k_0^2 - k_j^2 + \lambda k_0 k_j^2~, \\
(p\oplus_\ell p')_0=p_0+p'_0 & \qq (k\oplus_\lambda k')_0=k_0+k'_0 \\
(p\oplus_\ell p')_j=p_j+p'_j+\ell p_0 p'_j & \qq  (k\oplus_\lambda k')_j=k_j+k'_j+\lambda k_0 k'_j
\label{ex2}
\end{align}
This ``$\kappa$-Poincaré scenario'' is of interest from the point of view of DSR and also in recent studies in the context of ``relative-locality momentum spaces''~\cite{Gubitosi:2013rna,AmelinoCamelia:2011bm}.

Translating the scenario into our notation convention, and introducing a new constant $\rho$ such that $\ell\equiv 1/\Lambda$, $\lambda\equiv\rho/\Lambda$ (that is, by definition, $\rho=\lambda/\ell$), we have that in this case
\begin{equation}
\alpha_2^a=\gamma_1^a=1 \qq \alpha_2^b=\gamma_1^b=\rho \qq \alpha_1^a=\alpha_1^b=\beta_1^a=\beta_1^b=\beta_2^a=\beta_2^b=\gamma_2^a=\gamma_2^b=\gamma_3^a=\gamma_3^b=0
\label{ex2coefs}
\end{equation}
One can check that Eq.~\eqref{eq:golden} is satisfied for both particle sectors.

Ref.~\cite{Amelino-Camelia2012} shows that a consistent way to compose momenta of different types of particles is
\begin{equation}
(p\oplus k)_j=p_j+k_j+\frac{\ell+\lambda}{2}\,p_0 k_j \qq (p\oplus k)_0=p_0+k_0\,.
\label{ex2mixing}
\end{equation}
This corresponds to 
\begin{equation}
\beta_1^{ab}=\beta_2^{ab}=0 \qq \gamma_1^{ab}=\frac{1+\rho}{2} \qq \gamma_2^{ab}=\gamma_3^{ab}=0\,,
\label{ex2mixingcoefs}
\end{equation}
so that once again $\gamma_1^{ab}=(\gamma^a_1+\gamma^b_1)/2$, and indeed Eqs.~\eqref{beta1ab} and \eqref{beta2ab} are satisfied.

Ref.~\cite{Amelino-Camelia2012} also gives consistent relativistic laws of action of boosts:
\begin{equation}
[N_{[p]}, p_0] =  p_1 \qq 
[N_{[p]}, p_1] =  p_0 + \ell p_0^2 + \frac{\ell}{2} p_1^2 \qq
[N_{[k]}, k_0] =  k_1 \qq 
[N_{[k]}, k_1] =  k_0 + \ell k_0^2 + \frac{\lambda}{2} k_1^2 
\label{ex2boosts1}
\end{equation}
in the one-particle sectors and 
\begin{equation}
N_{[p \oplus_{\ell} p^\prime]} =N_{[p]}+N_{[p^\prime]}+\ell p_0 N_{[p']} \qq
N_{[k \oplus_{\lambda} k^\prime]} =N_{[k]}+N_{[k^\prime]}+\lambda k_0 N_{[k']} \qq
N_{[p \oplus  k]} =N_{[p]}+N_{[k]}+\frac{\ell+\lambda}{2}\,p_0 N_{[k]}
\label{ex2boosts2}
\end{equation}
in the two-particle sectors.
This corresponds to 
\begin{align}
\lambda_1^a=\lambda_1^b=0 \qq \lambda_2^a=1 \qq \lambda_2^b&=\rho \qq \lambda_3^a=-\frac{1}{2} \qq \lambda_3^b=-\frac{\rho}{2} \\
\eta_1^{La}=\eta_2^{Lb}=\eta_2^{Ra}=\eta_2^{Rb}=\eta_1^{Lab}=\eta_2^{Lab}=\eta_2^{Rab}=0 \qq \eta_1^{Ra}&=1 \qq 
\eta_1^{Rb}=\rho \qq \eta_1^{Rab}=\frac{1+\rho}{2}
\label{ex2boostscoefs}
\end{align}
[it is assumed again that the coefficients $(ba)$ are equal to the $(ab)$] and one can check that this choice indeed satisfies Eqs.~\eqref{MCL-GLT}, \eqref{etasimab}, \eqref{etaantisim1} and \eqref{etaantisim2}.

In summary, previous examples of non-universal kinematics are particular (and rather simple) choices of the generic coefficients presented here, and the long calculations to show their consistency with a relativistic theory, together with the extraction of the appropriate action of boosts become a trivial check of the consistency formulas presented in this work.  

\section{Physical processes}

In the rest of the paper we will apply the obtained conditions that the RP imposes in non-universal kinematics to specific physical situations, such as the generation of thresholds in particle decays and the ultrarelativistic limit of scattering processes with two-body final states. We will see that the presence of a RP has consequences for the modified kinematics in both cases which are qualitatively different from those in the Lorentz violation case. We will then consider a simple but interesting physical process: the elastic scattering of two particles in a BRK scenario in which one of the particles obeys to SR kinematics, while the other has a modified kinematics. A particular example of this process was considered in Ref.~\cite{Amelino-Camelia2012} in relation with a possible solution of the soccer-ball problem. We will re-examine the conclusions and conjectures presented there in the light of the general results of the present work.

\subsection{Thresholds in two-body particle decays}

The simplest physical process one can study is the decay of a particle $A$ into two other particles $C$ and $D$, $A\to C+D$. In general, a modified kinematics may produce thresholds so that a decay which is kinematically allowed (forbidden) in SR can become forbidden (allowed) at a certain energy. This is not surprising, since the balance between energy and momentum changes when there are modified dispersion relations and/or conservation laws. However, if the new kinematics is consistent with a relativity principle, then it cannot produce thresholds in a particle decay, since two observers could disagree whether the energy of the decaying particle is above or below the threshold, giving different physical predictions. This should be explicitly seen when the modified kinematics satisfy the consistency equations derived in this work. We will now show that this is indeed the case.

Let us consider the kinematics of the decay $A\to C+D$, where $A$ is a particle of type $(a)$ with four-momentum $k$, $C$ is a particle of type $(c)$ with four-momentum $p$, and $D$ is a particle of type $(d)$ with momentum $q$. The conservation law of this\footnote{\label{footnotechannel}In fact it turns out that this is only one of the 12 different conservation laws that are possible for this process. They correspond to different reordering of momenta and the use of antipodes for the in-going or out-going particles. For a complete discussion on this issue see Ref.~\cite{Carmona:2014lqa}.}
 process is
\begin{equation}
\hat{k}\oplus p \oplus q=0,
\label{eq:newconslaw}
\end{equation}
where $\hat{k}$ is the \emph{antipode} of $k$, that is, the four-vector that satisfies $\hat{k}\oplus k=k\oplus\hat{k}=0$. This modified kinematics was studied in general in Ref.~\cite{Carmona:2014lqa}. Using the conservation law to express the four-momentum $k$ as a function of $p$, $q$ and the dispersion relations of $C$ and $D$ to express the zero component of the four-momentum $p$, $q$ as a function of the modulus of the corresponding $\vec{p}$ and $\vec{q}$ vectors, one gets a relation between these vectors as a consequence of the dispersion relation of particle $A$. This equation was derived in Ref.~\cite{Carmona:2014lqa}, giving
\begin{equation}
2 E_p E_q - 2\vec{p}\cdot\vec{q}-m_a^2+m_c^2+m_d^2=O_3\,,
\label{eq:master}
\end{equation}
where $O_3$ contains all the terms which are proportional to $1/\Lambda$ coming from the modified kinematics, $m_a$, $m_c$ and $m_d$ are respectively the masses of particles $A$, $C$ and $D$ appearing in their respective modified dispersion relations [the variable $m$ appearing in Eq.~\eqref{eq:MDR}], and we have defined the variables 
\begin{equation}
E_p\equiv \sqrt{\vec{p}^2+m_c^2} \quad , \quad E_q\equiv \sqrt{\vec{q}^2+m_d^2}\quad \,.
\label{eq:SRenergies}
\end{equation}

Eq.~(15) of Ref.~\cite{Carmona:2014lqa} contains the expression of $O_3$, that we reproduce here:
\begin{align}
O_3&=\frac{E_p+E_q}{\Lambda}\left\{(\alpha_1^c+\alpha_2^c)E_p^2+(\alpha_1^d+\alpha_2^d)E_q^2+
(\hat{\alpha}_1^a+\hat{\alpha}_2^a) (E_p+E_q)^2 + 2 (\beta_1^{ac}+\beta_2^{ac}-\gamma_1^{ac}-\gamma_2^{ac})(E_p+E_q)E_p \right. \nonumber \\ 
& \left. + 2 (\beta_1^{ad}+\beta_2^{ad}-\gamma_1^{ad}-\gamma_2^{ad}) (E_p+E_q)E_q-2 (\beta_1^{cd}+\beta_2^{cd}-\gamma_1^{cd}-\gamma_2^{cd}) E_p E_q \right\} + \mathcal{O}\left(\frac{Em^2}{\Lambda}\right)\,,
\label{eq:masteroperator-MCL}
\end{align}
where in the last term $E$ stands for $E_p$ or $E_q$, and $m^2$ represents a squared mass or a combination of squared masses, so that this term represents in fact a sum of terms which are sub-dominant with respect to those which are explicitly written in the previous expression, in the ultrarelativistic limit $E_p^2\gg m_c^2$, $E_q^2\gg m_d^2$. On the other hand, $\hat{\alpha}^a_1$, $\hat{\alpha}^a_2$ are the coefficientes in the MDR of $\hat{k}$, the antipode of the momentum of the decaying particle. It can be easily shown (see Eqs.~(16)-(18) of Ref.~\cite{Carmona:2014lqa}) that 
\begin{equation}
\hat{\alpha}_1^a=-\alpha_1^a-2\beta_1^a\,, \quad \quad \hat{\alpha}_2^a=-\alpha_2^a-2(\beta_2^a-\gamma_1^a-\gamma_2^a)\,.
\label{eq:coefMDRantipode}
\end{equation}

It is convenient to rewrite the expression of $O_3$ in the form
\begin{equation}
O_3=\frac{(E_p+E_q)}{\Lambda}\left[\xi^{ac} (E_p+E_q)E_p + \xi^{ad} (E_p+E_q) E_q - \xi^{cd} E_pE_q\right]+ 
\mathcal{O}\left(\frac{Em^2}{\Lambda}\right)\,,
\label{eq:O3}
\end{equation}
with
\begin{align}
\xi^{ac} &= 2(\beta_1^{ac}+\beta_2^{ac}-\gamma_1^{ac}-\gamma_2^{ac})+(\hat{\alpha}_1^a+\hat{\alpha}_2^a)+(\alpha_1^c+\alpha_2^c) \\
\xi^{ad} &= 2(\beta_1^{ad}+\beta_2^{ad}-\gamma_1^{ad}-\gamma_2^{ad})+(\hat{\alpha}_1^a+\hat{\alpha}_2^a)+(\alpha_1^d+\alpha_2^d) \\
\xi^{cd} &= 2(\beta_1^{cd}+\beta_2^{cd}-\gamma_1^{cd}-\gamma_2^{cd})+(\alpha_1^c+\alpha_2^c)+(\alpha_1^d+\alpha_2^d)
\label{eq:xi} \,.
\end{align}

The modifications in the kinematics are important when the right-hand side of Eq.~\eqref{eq:master} is of the order of the left-hand side of that equation, that is, when $O_3 \sim m^2$. It is then immediate to note that sub-dominant terms are not useful to predict an energy threshold since this would mean
\begin{equation}
\frac{E_{th}m^2}{\Lambda}\sim m^2 \qq \Rightarrow \qq E_{th} \sim \Lambda\,,
\label{eq:threshold2}
\end{equation}
but all our formalism is valid as long as $E\ll\Lambda$. Therefore, in case there was a threshold, it should come from the dominant terms, which are cubic in energies, so that
\begin{equation}
\frac{E_{th}^3}{\Lambda}\sim m^2 \qq \Rightarrow \qq E_{th}\sim \left(m^2 \Lambda\right)^{1/3}\,.
\label{eq:threshold1}
\end{equation}  
However, when the ``golden rules''~\eqref{eq:golden} are combined with Eq.~\eqref{eq:coefMDRantipode} one finds that
\begin{equation}
\hat{\alpha}_1^a=\alpha_1^a=-\beta_1^a \quad \quad\quad\hat{\alpha}_2^a=\alpha_2^a=\gamma_1^a+\gamma_2^a-\beta_2^a\,,
\label{eq:goldenantipode}
\end{equation}
so that the antipode of a momentum satisfies the same MDR as the momentum as a consequence of the RP, and the three coefficients $\xi^{ac},\xi^{ad},\xi^{cd}$ are zero as a consequence of the relations~\eqref{eq:golden}, \eqref{eq:betaab} and \eqref{eq:gammaab} for the parameters of the MCL implied by the RP. Then, all the dominant terms in Eq.~\eqref{eq:O3} disappear in the case of a modified kinematics compatible with the RP. 

As a conclusion, our formalism correctly predicts the absence of thresholds in particle decays when a relativity principle is present. For an analysis in the case of a kinematics without such a restriction the reader is referred to Ref.~\cite{Carmona:2014lqa}. 

\subsection{Ultrarelativisic limit of two-body final state scattering processes}

Another common physical process is $2\to 2$ scattering, the collision of a particle $A$ [of type $(a)$] with four-momentum $k$ with a particle $B$ [of type $(b)$] with four-momentum $l$, giving a particle $C$ [of type $(c)$] with four-momentum $p$ together with a particle $D$ [of type $(d)$] with four-momentum $q$, that is, the process $A+B\to C+D$.

In this case the conservation law is\footnote{Once again (see the previous footnote) this is only one of 48 possible conservation laws for this process.}
\begin{equation}
\hat{k}\oplus \hat{l}\oplus p \oplus q=0\,,
\label{eq:newconslawscatt}
\end{equation}
where $\hat{k}$ ($\hat{l}$) is the antipode of the four-momentum $k$ ($l$). Proceeding as in the case of the decay, one can use the previous conservation law to express the four-momentum $k$ in terms of the momenta $\vec{l}$, $\vec{p}$ and $\vec{q}$ (the zero component of $l$, $p$, $q$ is written as a function of the corresponding vector by making use of the dispersion relation of each particle) and then the dispersion relation of particle $A$ leads to an equation for the three momenta $\vec{l}$, $\vec{p}$ and $\vec{q}$. The result can be expressed in the form
\begin{equation}
2(E_p E_q-\vec{p}\cdot\vec{q})-2(E_l E_p-\vec{l}\cdot\vec{p})-2(E_l E_q-\vec{l}\cdot\vec{q})+m_b^2+m_c^2+m_d^2-m_a^2=O_4\,,
\label{eq:master2}
\end{equation}
where again the $m_\alpha$ are the masses that appear in the dispersion relation of the particles, $E_l=\sqrt{\vec{l}^2+m_b^2}$ is added to the definitions given in Eq.~\eqref{eq:SRenergies}, and all the correction to the kinematics of this process in SR is contained in $O_4$, which is proportional to $1/\Lambda$ (at the order we are working throughout this work). After a long calculation, the result for $O_4$ is
\begin{equation}
\begin{split}
O_4 =\frac{(E_p+E_q-E_l)}{\Lambda}&\, [\xi^{ac} (E_p+E_q-E_l)E_p + \xi^{ad} (E_p+E_q-E_l) E_q - \xi^{ab} (E_p+E_q-E_l) E_l  \\
& - \xi^{cd} E_pE_q + \xi^{bc}E_lE_p + \xi^{bd}E_lE_q] +\mathcal{O}\left(\frac{Em^2}{\Lambda}\right)\,,
\end{split}
\label{eq:O4}
\end{equation}
where, together with $\xi^{ac}, \xi^{ad}, \xi^{cd}$ defined in Eq.\eqref{eq:xi}, one has
\begin{align}
\xi^{ab} &= 2(\beta_1^{ab}+\beta_2^{ab}-\gamma_1^{ab}-\gamma_2^{ab})+(\hat{\alpha}_1^a+\hat{\alpha}_2^a)+
(\hat{\alpha}_1^b+\hat{\alpha}_2^b) \\
\xi^{bc} &= 2(\beta_1^{bc}+\beta_2^{bc}-\gamma_1^{bc}-\gamma_2^{bc})+(\hat{\alpha}_1^b+\hat{\alpha}_2^b)+(\alpha_1^c+\alpha_2^c) \\
\xi^{bd} &= 2(\beta_1^{bd}+\beta_2^{bd}-\gamma_1^{bd}-\gamma_2^{bd})+(\hat{\alpha}_1^b+\hat{\alpha}_2^b)+(\alpha_1^d+\alpha_2^d)
\label{eq:xi-2} \,.
\end{align}
As a consistency check, one can note that Eq.~\eqref{eq:O3} can be obtained from Eq.~\eqref{eq:O4} by taking $E_l=0$.

Once again, we see that all terms which are dominant in the ultrarelativistic limit in Eq.~\eqref{eq:O4} (which are of course of order $E^3/\Lambda$) contain combinations of parameters that are zero in the case of a kinematics compatible with a relativity principle. This confirms the suppression observed in (universal) DSR theories with respect to kinematic consequences of a Lorentz violation~\cite{Carmona:2010ze} and extends it to the non-universal case.

We also note that the expression Eq.~\eqref{eq:O4}, here obtained for the first time, can be very useful in future phenomenological studies in the case of Lorentz violation (without a relativity between observers).    

\subsection{Example of a process with a bipartite relativistic kinematics}

Beyond the vanishing of the dominant terms in the ultrarelativistic limit as shown in the previous subsections, the study of the implications of a non-universal departure from SR kinematics compatible with the RP requires a case by case analysis of the correction to the corresponding kinematic equation. We will restrict ourselves to the simple example of the elastic scattering 
$A(k) + B(l) \to A(p) + B(q)$ in which $B$ is a particle of type $b$, which satisfies the SR kinematics 
\begin{equation}
\alpha_i^b =\beta_i^b = \gamma_i^b = 0 \,,
\end{equation} 
and $A$ is a particle of type $a$, with a kinematics with parameters $\alpha_i$, $\beta_i$, $\gamma_i$ in the modified dispersion relation and composition laws. We will also consider the simplest choice compatible with the RP for the parameters in the modified composition law between momenta of particles of the types $a$  and $b$
\begin{equation}
\beta_i^{ab} = \beta_i^{ba} = \frac{\beta_i}{2} {\hskip 1cm}
\gamma_i^{ab} = \gamma_i^{ba} =\frac{\gamma_i}{2} \,.
\end{equation}

Since there is a relativity principle, we can analyze the process in the system of reference in which particle $B$ is at rest, $l_\mu = (M, \vec{0})$. Then Eq.~\eqref{eq:master2} reduces to
\begin{equation}
2(E_p E_q - \vec{p}\cdot\vec{q})-2M(E_p+E_q)+2M^2=O_4\,,
\label{eq:MKE-elastic}
\end{equation}
and $O_4$ can be computed taking into account the conservation law~\eqref{eq:newconslawscatt}, together with the modified composition laws and dispersion relations of particles $A$ and $B$. A long calculation gives the result
\begin{equation}
O_4=-\frac{\gamma_1}{\Lambda}M\vec{p}\cdot(\vec{p}+\vec{q})+\frac{\gamma_1}{\Lambda}E_p\,\vec{q}\cdot(\vec{p}+\vec{q})+\frac{\gamma_2}{\Lambda}E_q\,\vec{p}\cdot(\vec{p}+\vec{q})+\frac{\gamma_1+\gamma_2}{\Lambda}(M-E_p-E_q)\,\vec{p}\cdot\vec{q}\,.
\label{eq:O4-ch1}
\end{equation}
Equation~\eqref{eq:MKE-elastic} can then be rewritten in the following form
\begin{equation} \label{eq:MKE-elastic-2}
\begin{split}
-2M(E_q-M)-2\vec{p}\cdot\vec{q}+2E_p(E_q-M)&=
\frac{\gamma_2-\gamma_1}{\Lambda}M\vec{p}\cdot(\vec{p}+\vec{q})
+\frac{\gamma_2}{\Lambda}(E_q-M)\,\vec{p}\cdot(\vec{p}+\vec{q})\\ &+
\frac{\gamma_1}{\Lambda}E_p\,\vec{q}\cdot(\vec{p}+\vec{q})-
\frac{\gamma_1+\gamma_2}{\Lambda}(E_q-M+E_p)\,\vec{p}\cdot\vec{q}\,.
\end{split}
\end{equation}

Ref.~\cite{Amelino-Camelia2012} considered this example of elastic scattering in connection with the soccer-ball problem, or the fact that a modified kinematics at a microscopic level should not translate into large corrections for macroscopic physics, which we certainly do not observe. The idea of Ref.~\cite{Amelino-Camelia2012} was that non-universality can implement a distinction between the kinematics of microscopic and macroscopic objects, so that macroscopic objects obey the kinematics of SR, while microscopic particles may have a modified kinematics. According to Ref.~\cite{Amelino-Camelia2012}, this would solve the soccer-ball problem if in the scattering of a microscopic particle with a macroscopic object there were not any pathological correction of the form $M/\Lambda$, which would be huge if $M$ were, for example, a soccer ball.

In a particular choice of modified relativistic kinematics, Ref.~\cite{Amelino-Camelia2012} showed indeed the absence of such pathological terms, and conjectured about the possibility that this were a generic result. To see whether it holds in our more general framework, let us take $M$ sufficiently large in Eq.~\eqref{eq:MKE-elastic-2} so that we can take $E_p\ll M$, $(E_q-M)\ll M$ in the terms which are proportional to $1/\Lambda$; then the equation of the process is
\begin{equation}
-2M(E_q-M)-2\vec{p}\cdot\vec{q}+2E_p(E_q-M)\approx
\frac{\gamma_2-\gamma_1}{\Lambda}M\vec{p}\cdot(\vec{p}+\vec{q})\,.
\end{equation}
Therefore the correction owing to the modification in the kinematics of the microscopic particle is of order $(M/\Lambda)$, where $M$ is the mass of the macroscopic object, instead of the naive correction of order $(E_p/\Lambda)$, and against the suggestions of Ref.~\cite{Amelino-Camelia2012}.  

One can trace out which is the difference between the previous work and ours. In Ref.~\cite{Amelino-Camelia2012} it is used a particular conservation law, which is not equivalent to~\eqref{eq:newconslawscatt} but to the conservation law
\begin{equation}
\hat{k}\oplus p \oplus \hat{l} \oplus q = 0 \,.
\label{eq:ACconslaw}
\end{equation}
This in fact corresponds to a different \emph{channel} of the process (we are using here the terminology of Ref.~\cite{Carmona:2014lqa}), that is, another possible conservation law (see footnote~\ref{footnotechannel}), which has the peculiarity that the momenta corresponding to the same kind of particle are adjoining. It turns out that this fact makes the pathological term absent. Indeed one can compute the equation of the process for this conservation law, which takes the form
\begin{equation} \label{eq:MKE-elastic-3}
-2M(E_q-M)-2\vec{p}\cdot\vec{q}+2E_p(E_q-M)=
\frac{\gamma_1}{\Lambda}E_p\,\vec{q}^2-
\frac{\gamma_2}{\Lambda}E_p \,\vec{p}\cdot\vec{q}+
\frac{\gamma_2}{\Lambda}(E_q-M)\,\vec{p}^2-
\frac{\gamma_1}{\Lambda}(E_q-M)\,\vec{p}\cdot\vec{q}\,.
\end{equation}
and indeed in this case the correction of order $(M/\Lambda)$ is gone.

If one interprets the presence of a pathological term of order $(M/\Lambda)$ as an inconsistency of the non-universal kinematics with the well-known macroscopic physics (as Ref.~\cite{Amelino-Camelia2012}) did), then this result indicates that not all the different channels (the non-equivalent conservation laws, corresponding to a different order of the momenta) are equally possible. This reasoning would lead to consider only those conservation laws in which the momenta corresponding to particles of the same type are composed together, and reject all the other possible combinations. This is rather noticeable: formal considerations would be allowing us to obtain some dynamical conclusions, even without any dynamical framework at our disposal.

However, we should be cautious and regard this as an open problem. It is not clear whether one can treat an elastic scattering between a microscopic particle and a macroscopic object in the same way as an scattering between microscopic particles, and in fact this problem is related to how the kinematics of composed systems depend on the kinematics of the components in a framework beyond special relativity, which is another way of stating the soccer-ball problem. A complete resolution of these issues might have as a bonus the possibility to amplify and observe certain effects of the modified kinematics in macroscopic systems (for other attempts to observe quantum gravity effects in macroscopic systems, see e.g.~\cite{Pikovski2012,PhysRevD.86.124040}).

\section{Conclusions}

In a kinematics beyond special relativity, the RP imposes consistency conditions between the modified dispersion relation, the modified composition law, and the generalized Lorentz boosts. We have extended a previous work showing these conditions for a universal kinematics to the case of a non-universal kinematics. The conditions presented here are a powerful tool in the exploration of non-universal kinematics consistent with a RP, since long calculations to show this consistency and to get the appropriate actions of boosts, such that those presented in Ref.~\cite{Amelino-Camelia2012}, become a trivial check of the consistency formulas presented here. 

We have shown that the common lore stating that the presence of a RP suppresses the consequences of a modified kinematics with respect to the case of Lorentz violation is a valid conclusion also in the non-universal case, by seeing explicitly that the dominant terms in the modified part of the equation that describes the kinematics of a $2\to 2$ process are zero in the relativistic case. This argument can be easily extended to more general processes.

Non-universal kinematics consistent with a RP could be of great, even fundamental, importance if special relativity were to be modified by quantum gravity effects in such a way that there is no preferred reference system, as it happened with the transition of Galilean to special-relativistic kinematics. There is no reason why these effects should be universal, and in fact non-universality could be the key to solve the soccer-ball problem, or, more generally, how kinematic corrections to elementary particles translate to composed systems.

We have considered the simplest type of non-universality, that of a bipartite relativistic kinematics, and see how previous examples of non-universality in the literature are in fact specific examples of it. 

We have analyzed the particular case of an elastic scattering between two particles which obey, respectively, the kinematics of SR and a modified kinematics, and shown that it could be relevant in the resolution of the soccer-ball problem and the possibility to amplify effects of the modified kinematics by using macroscopic systems.

In a different line of thought, and forgetting for the moment about quantum gravity effects, a possible application of a bipartite relativistic kinematics could be a situation where the known particles obey the kinematics of special relativity, but there is an unknown, not-yet-discovered, dark sector, whose kinematics is different from that of special relativity. If these were the case, experiments trying to detect this sector (dark matter experiments, axion or WISPs searches) could miss some fundamental clue. Though we acknowledge that this is a very speculative scenario, it could be an interesting line of research for future work.  

\section*{Acknowledgments}
This work is supported by the Spanish MINECO (FPA2012-35453) and DGIID-DGA (grant 2011-E24/2).

\end{document}